\documentclass[12pt]{article}
\pdfoutput=1

\usepackage{fullpage}
\usepackage{amsmath, amssymb, hyperref}

\newcommand{\ud}{\text{d}}

\begin{document}

\title{Non-Abelian R-symmetries in $\mathcal{N}=1$ supersymmetry}
\author{James Brister\textsuperscript{\dag},
        Longjie Ran\textsuperscript{\ddag},
        Zheng Sun\textsuperscript{*}\\
        \normalsize\textit{College of Physics, Sichuan University, Chengdu 610064, P.\ R.\ China}\\
        \normalsize\textit{E-mail:}
                   \parbox[t]{26em}{
                   \textsuperscript{\dag}\texttt{jbrister@scu.edu.cn,}
                   \textsuperscript{\ddag}\texttt{2019222020004@stu.scu.edu.cn,}
                   \textsuperscript{*}\texttt{sun\_ctp@scu.edu.cn}}
       }
\date{}
\maketitle

\begin{abstract}
We investigate non-Abelian R-symmetries in $\mathcal{N}=1$ supersymmetric theory, where fields may transform under the R-symmetry in representations with dimension higher than one.  While a continuous non-Abelian R-symmetry can always be decomposed to a $U(1)$ R-symmetry and non-R symmetries, there are non-trivial discrete non-Abelian R-symmetries that do not admit such a decomposition, and effective R-charges cannot be defined in such models.  Previous results on sufficient conditions for R-symmetric supersymmetric vacua in Wess-Zumino models still hold, and do not depend on fields in representations of dimension greater than one.  However, fields in higher-dimensional representations enter the sufficient conditions for supersymmetric vacua that break R-symmetry, but it is difficult to identify the independent variables which can be used to solve the F-flatness equation in this case, unless other conditions are fulfilled.  We present examples with discrete non-Abelian R-symmetries of the lowest order in this case.
\end{abstract}

\section{Introduction}

The Nelson-Seiberg theorem relates R-symmetries to F-term supersymmetry (SUSY) breaking in generic Wess-Zumino models~\cite{Nelson:1993nf}.  Several refinements to the theorem have recently been made~\cite{Sun:2011fq, Kang:2012fn, Sun:2019bnd, Li:2020wdk, Amariti:2020lvx, Sun:2021svm, Li:2021ydn, Brister:2022rrz, Nakayama:2023eax}, and subsequently been verified in a large set of searched models~\cite{Brister:2022vsz, Sun:2022xdl}.  These results have all been derived in the context of Abelian (and generally continuous) R-symmetries.  However, it is also possible to have Non-Abelian discrete R-symmetries in $\mathcal{N}=1$ supersymmetry, which may be useful for phenomenology model building~\cite{Chen:2013dpa}.  In this paper, we show that non-trivial non-Abelian R-symmetries indeed lead to different generic models than their Abelian counterparts.  We derive the generalizations to the Nelson-Seiberg theorem in this case, and demonstrate that generic models can be built that exhibit all possible combinations of R-symmetry and SUSY breaking.

The structure of the paper is as follows.  Section \ref{setup} introduces generic Wess-Zumino models and R-symmetries.  Section \ref{prev} reviews the previously established results on this topic.  Section \ref{nonequiv} establishes that non-Abelian R-symmetries lead to models that are not equivalent to any with an Abelian R-symmetry.  Section \ref{vacua} derives sufficient conditions for SUSY and R-symmetry breaking.  Section \ref{groups} discusses, and resolves, some difficulties in model bulding and construct example models that exhibit all of the possible symmetry breaking behaviors found in the previous sections.

\section{Set-up}\label{setup}

We work in an $\mathcal{N}=1$ supersymmetric Wess-Zumino model: we have chiral superfields $\Phi_i$ that form a scalar superpotential $W(\Phi_i)$, which appears in the superspace action as
\begin{equation}
S_W = \int \ud^4 x \, \mathcal{L}_W
    = -\int \ud^4 x \,
           \left ( \int \ud \theta^\alpha \ud \theta_\alpha \, W + \text{c.c.} \right ).
\end{equation}
We will suppress the fermion indices $\alpha$ for the remainder of this article. For simplicity, we assume that this superpotential is renormalizable, being a polynomial in the superfields of up to cubic order.\\

The potential for the scalar components $\phi_i$ of the superfields, is given by
\begin{equation}
V(\phi) = (\partial_i W)(\partial_j W)^*K_{ij}
\end{equation}
Where $K_{ij}$ is the K\"ahler metric defined as the matrix inverse of the derivative of the K\"ahler potential with respect to $\Phi$ and $\Phi^*$, which we may take to be trivial (so that $K_{ij}=\delta_{ij}$) without affecting any of the results we derive in this paper.

A vacuum (that is, a minimum of $V$) will preserve supersymmetry only if the field  vacuum expectations (v.e.v s) obey the ``F-term'' flatness equations
 \begin{equation}
\partial_i W =0~\forall i~.
\end{equation}
It is clear that any solutions to $\partial_i W =0~\forall i$ must be global minima of $V$; to determine whether supersymmetry is broken through F-terms we need only determine whether there is a solution to $\partial_i W =0~\forall i$.

\subsection{R-symmetries}

An R-symmetry is a symmetry that acts non-trivially on $\theta$, the superspace grassman-number coordinates.  In $\mathcal{N}=1$ supersymmetry, this means that the R-symmetry must act on $\theta$ as a one-dimensional representation.  Assuming unitarity, the R-symmetry must thus act either as a discrete $\mathbb{Z}_n$:  $\theta \to \omega \theta$, where $\omega$ is a complex number such that $\omega^n = 1$ for some integer $n\ge 3$; or else as a continuous $U(1)$: $\theta \to e^{i \alpha}\theta, ~ \alpha \in [0, 2\pi)$.  Note that a discrete R-symmetry with $n=2$, or R-parity, is possible.  But it does not change the superpotential, and the transformation of $\theta$ can be absorbed in a Lorentz rotation.  Therefore, we consider discrete R-symmetry with $n \ge 3$ in the following work.

We focus here on the case in which this R-symmetry is a representation of a larger, non-Abelian group $G$.  The continuous case is trivial: if we take $G$ to be a compact Lie group, then by Levi decomposition~\cite{Barut1977}, we must have $G= H \times U(1)_R$, where $H$ is some non-R flavor symmetry.  This reduces to the Abelian $U(1)_R$ case studied in previous literature~\cite{Nelson:1993nf, Sun:2011fq, Kang:2012fn, Sun:2019bnd, Li:2020wdk, Amariti:2020lvx, Sun:2021svm, Li:2021ydn, Brister:2022rrz, Nakayama:2023eax, Brister:2022vsz, Sun:2022xdl}.  We thus assume a discrete R-symmetry, which we take to be part of a larger non-abelian symmetry group $G$ which does not decompose as $H \times \mathbb{Z}_n$, where $\mathbb{Z}_n$ generates the R-symmetry.

\subsection{Linear representations and notation}

In order to act as an R-symmetry, $G$ must have a non-trivial one-dimensional representation under which $\theta$ may transform. We shall denote this representation as $1_1$. That is, we have an injective homomorphism
\begin{equation}
1_1: G \to \mathbb{Z}_n
\end{equation}
By taking powers of this representation, we may then obtain another $n-1$ representations $1_2, 1_3, \ldots, 1_{n-1}, 1_0$ with
\begin{equation}
1_a \otimes 1_b = 1_{a+b \text{ mod } n}
\end{equation}
where $1_0$ is of course the trivial representation.

If $\Gamma$ is an irreducible representation of $G$, then so is $\Gamma \otimes 1_a$ . We therefore partition the irreducible representations of $G$ into sets $\{K_a \}$, where the index $a$ is chosen such that
\begin{equation}\label{K}
K_a \otimes 1_b = K_{a+b \text{ mod } q}
\end{equation}
for some integer $q$ that divides $n$.

For the groups we shall consider in this paper, there is only one such set of $K_a$ for each dimension of irreducible representation that the group has. We can thus denote these representations as $1_a, 2_b, 3_c$ etc.

\subsection{Genericity}

As we seek results related to the Nelson-Seiberg theorem, we assume the superpotential to be generic. That is, we assume the superpotential is a linear combination of all terms allowed by the symmetries of the theory and renormalizability, and that there are no relations between their coefficients.  More precisely, we mean generic in the sense of~\cite{Brister:2021xca}.

As $\theta$ (by assumption) transforms in the $1_1$ representation of $G$, the superpotential must transform in the $1_2$ representation.  That is: $W$ is a sum, with generic coefficients, of all the polynomial combinations of the $\Phi_i$, up to cubic order, that transform in the $1_2$ representation.  An important consequence is that only a field that transforms in the $1_2$ representation can appear as a linear term $a\Phi_i$ in $W$.

\section{Previous results}\label{prev}

In \cite{Nelson:1993nf} it was noted that, in the absence of an R-symmetry, the system of equations $\partial_i W =0$ is $n$ equations in $n$ unknowns, and thus will generically have a solution.  A non-R symmetry cannot change this, as the number of independent equations will be decreased by the same number as variables that are removed.  An R-symmetry, however, produces a constraint on the form of the superpotential, which removes some variables from certain F-term equations, and so a generic superpotential may possibly not have a supersymmetric vacuum.  Thus an R-symmetry is a necessary condition for SUSY breaking in a generic model.  It is also argued that a broken R-symmetry is a sufficient condition for SUSY breaking, but the argument rests on an additional assumption of genericity on the form of the superpotential, which can be easily violated by renormalizability or other restrictions to the superpotential.  The exceptions can be seen in~\cite{Sun:2019bnd, Amariti:2020lvx, Sun:2021svm, Li:2021ydn, Brister:2022rrz, Nakayama:2023eax}, and we do not make such an additional assumption here.

Recent works~\cite{Kang:2012fn, Li:2020wdk} have attempted to refine these results further: finding necessary and sufficient conditions for SUSY breaking, expressed in terms of the R-charges of various fields that appear in a generic superpotential.  In particular, it was shown that a necessary condition for SUSY breaking is having a greater number of fields that transform in the same representation as $W$ itself than the number of fields that are invariant under the $R$-symmetry.

In~\cite{Brister:2021xca}, it was shown that for a continuous R-symmetry, there are very strong constraints on the form of a generic solution to $\partial_i W =0$: the superpotential $W$ must vanish termwise, a refinement on the well-established result that the superpotential must vanish overall~\cite{Kappl:2008ie, Dine:2009sw}.  As we are working with discrete R-symmetries, these results will not apply.  Indeed, in section \ref{nonvanishing} we will see a superpotential that does not vanish at all at a SUSY vacuum.

\section{Non-abelian R-symmetries are not equivalent to an abelian R-symmetry}\label{nonequiv}

Before we continue, we ought to establish that non-abelian R-symmetries really do allow different superpotentials than an abelian R-symmetry, even one with additional non-R symmetries.

Checking the list of finite groups of small order~\cite{Grouplist:website}, we note that, in order to have a one-dimensional representation of order at least 3, a non-abelian R-symmetry group must be of order at least 12. There are two groups of order 12 with appropriate one-dimensional representations: the group $Q_6$ given below, with a representation of order 4, and the alternating group on four elements, $A_4$, with a representation of order 3.

For simplicity, let the R-symmetry group be 
\begin{equation}
Q_6 = \mathbb{Z}_3 \rtimes \mathbb{Z}_4 = \langle a, b | a^6 = e, b^2=a^3, bab^{-1} = a^{-1}\rangle~.
\end{equation}
We assume $\theta$ transforms under the representation $1_1$, where $1_1(b) = i, 1_1(a)=-1 $.

Let $X, B_1, B_2, C_1$, and $C_2$ be (the scalar parts of) chiral superfields. Assume that $X$ transforms in the $1_2$ representation like $W$, and that the $B_i$ and $C_i$ both transform under the two-dimensional irreducible representation $2_1$, given by
\begin{equation}
2_1(a) = \begin{pmatrix}
e^{\frac{2\pi i}{6}}&0\\0&e^{\frac{-2\pi i}{6}}
\end{pmatrix}, ~
2_1(b) = \begin{pmatrix}
0 & 1 \\ -1 & 0
\end{pmatrix}
\end{equation}
Decomposing the tensor product $2_1 \otimes 2_1$ into irreducible representations, we find that $B_1 C_2 - B_2 C_1$ transforms as $1_0$ and $B_1 C_2 + B_2 C_1$ transforms as $1_2$, as do $B_1 B_2$ and $C_1 C_2$. The remaining components of $B_i C_j, B_i B_j$ and $C_iC_j$ transform as the other two-dimensional irreducible representation of $Q_6$: $2_2 = 2_1 \otimes 1_1$; there are no products of three fields that transform as a $1_2$ other than $X^3$.

A generic, renormalizable superpotential formed from these fields is thus 
\begin{equation}\label{Q6}
W =  \alpha X + \beta X(B_1 C_2 - B_2 C_1) +\gamma (B_1 C_2 + B_2 C_1) + \delta B_1 B_2 + \epsilon C_1 C_2+ \zeta X^3~,
\end{equation}
for generic coefficients $\alpha, \beta, \gamma, \delta, \epsilon, \zeta$.  Note that there is no way to consistently assign a scalar R-charge to the fields above.  If $B_1, B_2, C_1$ and $C_2$ all transformed in one-dimensional representations, it would be impossible for $(B_1 C_2 - B_2 C_1)$ and $(B_1 C_2 + B_2 C_1)$ to have different charges.  Similarly, we could not replace the quadratic $BC$ terms above with scalar-charged linear fields, as genericity would demand additional terms, and we would lose the relations between some of the terms.  Thus non-abelian R-symmetries indeed permit superpotentials that would not be possible under an abelian discrete R-symmetry.

\section{Results on SUSY vacua}\label{vacua} 

\subsection{Conditions for the existence of  SUSY vacua}

Following~\cite{Kang:2012fn}, we note that in the presence of an R-symmetry, it is only the existence of fields that transform as $1_2$ that allows for the possibility of SUSY breaking: without such fields, every term in the superpotential is at least quadratic, and so $\partial_i W = 0$ can always be satisfied by setting all the $\Phi_i$ to zero.

We shall denote the fields that transform under the $1_2$ representation as $X_i$. A generic superpotential can then be written schematically as
\begin{eqnarray}
W &=& X_i( \alpha_i + \beta_{ij}Y_j +\beta'_{i(jk)}Y_jY_k+ \gamma_{ia}c^a_{lm}B_l B_m ) +\gamma'_{(ij)k}X_iX_jA_k\nonumber\\
 &+& \text{quadratic and cubic terms that transform as } 1_2 \text{ not involving } X_i
\end{eqnarray}
where $\alpha, \beta, \beta', \gamma, \gamma'$ are generic coefficients, $Y_j$ are fields transforming as $1_0$, $A_k$ are fields transforming as $1_{-2 \text{ mod } n}$.  The $B_l$ are fields transforming in other non-trivial representations of the R-symmetry group, and the $c^a_{lm}$ are a linearly independent set of matrices, indexed by $a$, such that $c^a_{lm}B_l B_m$ transforms as $1_0$.  For now, we do not rule out the case $n=4$, in which we should identify the $X_i$ and $A_k$ fields.

While $\partial_i W=0$ as a whole is still $n$ equations in $n$ variables, consider the subset of those equations given by the $X_i$ derivatives:
\begin{equation}\label{XF}
\partial_{X_i} W = \alpha_i + \beta_{ij}Y_j +\beta'_{i(jk)}Y_jY_k+ \gamma_{ia}c^a_{lm}B_l B_m +2\gamma'_{(ij)k}X_jA_k=0~;
\end{equation}
This is how SUSY may be broken generically in the presence of an R-symmetry: only a certain subset of fields may appear in the F-term equations that come from fields that transform under a particular representation of the R-symmetry.  In particular, only the F-term equations for the $X_i$ have a constant term, coming from $\alpha_i X_i$; in the absence of such terms, one may always find a solution by setting the vacuum expectations to zero.

We will thus focus on the $X_i$ F-terms, as given in equation \ref{XF}.  These are a number of equations equal to the number of $X$ fields, $N_X$, in the number of independent variables $N_v$ from the $Y_j$ , $X_jA_k$ and $c^a_{lm}B_l B_m$ terms.  Clearly the $Y$ fields contribute one independent variable each, as they can appear linearly, and so $N_Y$, the number of $Y_k$ fields, contributes to $N_v$.  If there is at least one field $A_i$, we have $N_{AX} = N_A + N_X -1$ independent variables from the products $X_j A_k$, where $N_A$ is the number of $A_i$ fields.  Terms coming from fields $B_i$ transforming in non-trivial representations of the $R$-symmetry group only appear as certain combinations of quadratic terms, and some calculation may be required to work out how many of these terms are independent.  We shall refer to this number as $N_Q$ (for quadratic), which will in general be less than the number of independent components of the $B_i$ fields.  Thus we have $N_v= N_Y + N_{AX} + N_Q$.

\subsubsection{Necessary conditions for SUSY vacua}

A necessary condition for a generic SUSY vacuum is that the $X_i$ F-term equations generically have a solution.  If $N_X > N_Y + N_{AX}+ N_Q$, then we know that these equations generically have no simultaneous solutions and so supersymmetry must be broken (if a vacuum exists at all).  Thus $N_Y + N_{AX}+ N_Q \ge N_X$ is a necessary condition for the generic existence of a SUSY vacuum.

If there are any fields $A_i$ transforming as $1_{-2 \text{ mod } n}$, this necessary condition becomes trivial, as $N_{AX}$ would be at least $N_X$. This differs from the equivalent case for continuous R-symmetries, in which we know that a generic SUSY vacuum must have $X_i=0$ \cite{Brister:2021xca}, and so such terms can be excluded from the count of independent variables.  We will therefore focus on models in which $N_A = 0$ and thus $ N_{AX}= 0$, such that our necessary condition becomes  $N_Y + N_Q \ge N_X$.

By making additional assumptions on the form of a solution to the F-term equations, one could also derive sufficient conditions for a SUSY vacuum.  As a simple example, if we assume that all fields vanish at a SUSY vacuum, we easily derive the sufficient condition $N_X =0$.

\subsection{R-symmetric SUSY vacua}

If $N_Y\ge N_X$ then we may be able to solve $\partial_{X_i} W=0$ with $Y_j$ fields alone.  The $Y_j$ can only appear in terms proportional to $X_i$ or in terms of the form $a_{jkl}Y_jC_kC_l$, where the product $a_{jkl}C_kC_l$ transforms as a $1_2$.  Thus, if $N_Y \ge N_x$, we always (generically) have a SUSY vacuum in which some of the $Y_j$ gain a nonzero v.e.v, and all other fields are set to zero.  In this SUSY vacuum, the R-symmetry is also unbroken.  $N_Y \ge N_X$ is thus a necessary condition for $R$-symmetry to be generically preserved at the SUSY vacuum, else we must be in the scenario described in the next section.

\subsection{R-breaking SUSY vacua} \label{nonvanishing}

If we have $0 < N_X - N_Y \le N_Q +N_{AX}$, any SUSY vacuum will give a non-zero v.e.v to some quadratic combination of fields that transform nontrivially under the R-symmetry, thus giving a vacuum where SUSY is preserved, but R-symmetry broken.  Note that if $N_Y \ge N_X$, but there are other fields that contribute to $N_v$, there may be a degeneracy between vacua that preserve the $R$-symmetry and those that break it.

As an example, the superpotential given in equation \ref{Q6} has a SUSY vacuum at
\begin{eqnarray}
X= \pm \sqrt{\frac{\gamma^2-\delta\epsilon}{\beta^2}},\nonumber\\
B_2 = -\frac{1}{\delta}(\beta X + \gamma)C_2,\nonumber\\
B_1 = -\frac{1}{\delta}(-\beta X + \gamma)C_1,\nonumber\\
B_2C_1 = \frac{(\alpha+ 3 \zeta X^2)(\beta X+\gamma)}{2\beta^2 X} ~.
\end{eqnarray}
Note that the vacuum has a continuous degeneracy $B_1 \to \lambda B_1,~ B_2 \to \lambda^{-1} B_2,~ C_1 \to \lambda C_1,~ C_2 \to \lambda^{-1} C_2$.  This is generic for $R$-symmetry breaking SUSY vacua, as products of fields gain a vacuum expectation.

Moreover, the superpotential $W$ is nonzero at the vacuum.  SUSY vacua with discrete R-symmetries do not obey the conditions found in~\cite{Brister:2021xca, Kappl:2008ie, Dine:2009sw}, as these conditions are derived from considering infinitesimal transformations by a continuous R-symmetry.

It is difficult to say much in general about models like this, as it is unclear without explicit calculation which of the fields in the model will obtain vacuum expectations at a SUSY vacuum.

\section{Model building and non-abelian R symmetry groups}\label{groups}

\subsection{A problem with groups of low order}

From section \ref{vacua}, it is clear that finding conditions for SUSY breaking depends on products of fields that transform under the $1_2$ representation and linear or quadratic products of fields that transform under the $1_0$ representation.

The number of independent $1_2$ fields, $N_X$ is easy to determine unambiguously, but the counting of the number \emph{independent} products of fields that can be used to solve the F-term flatness equations is made significantly more complicated if, as in equation \ref{Q6}, the same fields can appear in both combinations that transform as $1_2$ and combinations that transform as $1_0$.  In fact, with that choice of group, and many others, this behavior is unavoidable.

Recall equation \ref{K}:
\begin{equation}
K_r \otimes 1_a = K_{r+a~\text{mod}~q}
\end{equation}
If, for some representation $\Gamma$, the tensor product $K_r \otimes \Gamma$ contains the irrep $1_a$, then as $K_r = K_{r+q} = 1_q \otimes K_r$, the product
\begin{equation}
K_r \otimes \Gamma = 1_q \otimes (K_r \otimes \Gamma)
\end{equation}
must also contain $1_{a+q}, 1_{a+2q}$ etc.

Thus, in order to have products of higher-dimensional representations that can contain $1_0$ but not $1_2$, we must have $q \ge 3$.  There is no group with this property with order less than 24.

\subsection{The smallest groups with $q \ge 3$}

There are two groups of order 24 with the required properties. We shall denote these as $G_1$ and $G_2$.  $G_1$ is the semi-direct product group  $\mathbb{Z}_3\rtimes \mathbb{Z}_8$, with presentation
\begin{equation}
G_1 = \langle a, b~|~ a^3 = b^8 = e,~ bab^{-1} = a^2 \rangle
\end{equation}
$G_2$ is  the ``binary tetrahedral group''\footnote{the double cover of the isometry group of a regular tetrahedron} $2T = SL_2(\mathbb{F}_3) = Q_8 \rtimes \mathbb{Z}_3$, with presentation
\begin{equation}
G_2=  \langle s,t~ |~ s^4 = t^3 = (st)^3 =e,~ s^2t =ts^2 \rangle~.
\end{equation}

For the purposes of model building, we will discuss the representations of both of these groups.

\subsection{Representation theory of $G_1$}

$G_1$ has 8 one dimensional representations $1_n,~ n=0,1,\ldots 7$, given by
\begin{equation}
1_n(a)=1,~ 1_n(b) = e^{\frac{i n\pi}{4}}
\end{equation}
and 4 two dimensional irreducible representations $2_m,~m=0,1,2,3$ given by
\begin{equation}
2_m(a) = \begin{pmatrix}
e^{\frac{i 2\pi}{3}}&0\\0&e^{\frac{i 4\pi}{3}}
\end{pmatrix}
,~
2_m(b) = \begin{pmatrix}
0&1\\i^m&0
\end{pmatrix}
\end{equation}
$2_1$ and $2_3$ are faithful representations of $G_1$.

\subsubsection{Tensor product decompositions}

We have 
\begin{eqnarray}
1_a \otimes 1_b = 1_{a+b~\text{mod}~ 8} \\
1_a \otimes 2_b = 2_{a+b~\text{mod}~ 4} \\
2_0 \otimes 2_0 = 1_0 \oplus 1_4 \oplus 2_0
\end{eqnarray}
The other tensor products of the $2_a$ can be inferred from the above\footnote{In the product of any two dimensional irreps, where $A_i$ transforms as a $2_k$, and $B_i$ as a $2_l$, it is two linear combinations of $A_1 B_2$ and $A_2 B_1$ that transform in the one-dimensional representations, as these are invariant under the action of $2_k(a)2_l(a)$}.  The two-dimensional representations have periodicity $q=4$ and, as we required above, there are tensor products of two representations that contain components that transform as $1_0$, without any that transform as $1_2$.

\subsection{Representation theory of $G_2$}

$G_2$ has seven inequivalent irreducible representations: three one-dimensional representations $1_a$, three faithful two-dimensional irreps $2_b$ and one three-dimensional irrep $3_0$. In appropriate co-ordinates, these are given by
\begin{eqnarray}
&1_a(s)& =1~,~1_a(t)= \omega^a,~~~a=0,1,2 \\
&2_b(s)& = \frac{-i}{\sqrt{3}}\begin{pmatrix}
1&\sqrt{2}\\\sqrt{2}&-1
\end{pmatrix}~,~
2_b(t) = \begin{pmatrix}
\omega^{b+2}&0\\0&\omega^{b+1}
\end{pmatrix}~~~b=0,1,2 \\
&3_0(s)& = \frac{1}{3}\begin{pmatrix}
-1&2&2\\2&-1&2\\2&2&-1
\end{pmatrix}~,~ 
3_0(t)= \begin{pmatrix}
1&0&0\\0&\omega&0\\0&0&\omega^2
\end{pmatrix}
\end{eqnarray}
where $\omega = e^{2\pi i /3}$

\subsubsection{Tensor product decompositions}

As expected, we have
\begin{eqnarray}
1_a \otimes 1_b &=& 1_{a+b \text{ mod }3}\\
1_a \otimes 2_b &=& 2_{a+b \text{ mod }3}\\
1_a \otimes 3_0 &=& 3_0
\end{eqnarray}
So that the two-dimensional representations have periodicity $q=3$.

Less trivially, we have
\begin{eqnarray}
2_0 \otimes 2_0 = 1_0 \oplus 3_0\\
3_0 \otimes 3_0 = 1_0 \oplus 1_1 \oplus 1_2 \oplus 3_0 \oplus 3_0\\
2_0 \otimes 3_0 = 2_0 \oplus 2_1 \oplus 2_2
\end{eqnarray}
products involving $2_1, 2_2$ are easily derived from the above\footnote{
We note that, given fields $(A_1, A_2)$ transforming as a $2_a$ and $(B_1, B_2)$ transforming as $2_b$, it is the antisymmetric combination $(A_1 B_2 - A_2 B_1)$ that transforms as a $1_{a+b}$, as it is invariant under the action of the generator $s$. The remaining components form the vector $(-A_1B_1, A_2B_2, \frac{1}{\sqrt{2}}(A_1B_2+A_2B_1))$ which transforms as a $3_0$, though not necessarily in the same basis as shown above.}

\subsection{Model building with $G_1$ and $G_2$}

Using these groups, we can easily build families of models that generically exhibit all the possible symmetry-breaking behaviors described in section \ref{vacua}.\\

As an example, let us take the R-symmetry group to be $G_1$, and let $X_a, Y_\alpha$ and $B_i^{(k)}$ be chiral superfields, where the $X_a$ transform in the $1_2$ representation (like $W$), the $Y_\alpha$ are uncharged and the fields $B_i^{(k)}$ $i=1,2$ transform in the $2_0$ representation.  The symmetric combinations $B_1^{(k)}B_2^{(l)} + B_2^{(k)}B_1^{(l)}$ are also uncharged, for any $k,l$ (including $k=l$), and there are no products of 2 or 3 $B$ fields in the $1_2$ representation.  A generic renormalizable potential is 
\begin{equation}
W= X_a\left(p_a(Y_\alpha) + c_{a(kl)}(B_1^{(k)}B_2^{(l)} + B_2^{(k)}B_1^{(l)})\right)~,
\end{equation}
where the $p_a$ are quadratic polynomials in the $Y_\alpha$ fields, and the $c_{a(kl)}$ are symmetric on their last two indices.

Attempting to solve $X_a=0, \partial_{X_a}W =0$, we see that we have $N_X$ equations to solve in a number of variables equal to $N_Y$, plus the number of independent components of $B_1^{(k)}B_2^{(l)} + B_2^{(k)}B_1^{(l)}$.  In this case, the latter number is easy to calculate: it is $2N_B -1$, where $N_B$ is the number of $B^{(k)}$ doublets.\\

For this superpotential, our neccesary condition for a SUSY vacuum is also sufficient\footnote{as we can solve the non-$X$ F-term equations by setting $X=0$.}, and so, by appropriately choosing the number of $X,~Y$ and $B$ fields, we can easily achieve any of the scenarios described in section \ref{vacua}: we generically have vacua that preserve both SUSY and R-symmetry if $N_Y \ge N_X$, vacua that preserve SUSY but break R-symmetry if $2N_B -1 \ge N_X -N_Y$, and no SUSY vacua at all if $N_X > N_Y + 2N_B - 1$.

\section{Conclusions}

In this paper, we have demonstrated that non-Abelian discrete R-symmetries can lead to genuinely different models.  We have demonstrated that suitable generalizations to the Nelson-Seiberg theorem still hold, with both broken and unbroken R-symmetries, and shown that it is possible to explicitly construct models that exhibit all of these behaviors.  This opens up unexplored avenues for Wess-Zumino type theories, with R-symmetries non-trivially mixed with non-Abelian flavor symmetries, which may lead to more tightly constrained models with rich phenomenology.  It is also worth to explore how to realize such non-Abelian R-symmetries from the geometry of the compact space in string phenomenology models.

\section*{Acknowledgement}

The authors thank Yan He and Zhengyi Li for helpful discussions.  This work is supported by the National Natural Science Foundation of China under the grant number 12205208 and 11305110.

\end{document}